\documentclass[lefteqn]{iopart}
\usepackage{graphicx}
\usepackage{iopams}
\newcommand{\url}{\tt}

\newcommand{\beq}{\begin{equation}}
\newcommand{\eeq}{\end{equation}}
\newcommand{\bea}{\begin{eqnarray}}
\newcommand{\eea}{\end{eqnarray}}

\begin{document}

\title[Charge and rotation rate]
{A universal constraint between charge and rotation rate for degenerate black holes surrounded by matter}
 \author{Marcus Ansorg$^1$ and Herbert Pfister$^2$}
 \address{$^1$ Max-Planck-Institut f\"ur Gravitationsphysik, Albert-Einstein-Institut, D-14476 Golm, Germany}
 \address{$^2$ Institut f\"ur Theoretische Physik, Universit\"at T\"ubingen, Auf der Morgenstelle 14, D-72076 T\"ubingen, Germany}
 \eads{\mailto{Marcus.Ansorg@aei.mpg.de},\quad \mailto{herbert.pfister@uni-tuebingen.de}}
 \date{\today}
\begin{abstract}
We consider stationary, axially and equatorially symmetric systems consisting of a central rotating and charged degenerate black hole and surrounding matter. We show that $a^2+Q^2=M^2$ always holds provided that a continuous sequence of
spacetimes can be identified, leading from the Kerr-Newman solution in electrovacuum to the solution in question. The quantity $a=J/M$ is the black hole's intrinsic angular momentum per unit mass, $Q$ its electric charge and $M$ the well known black hole mass parameter introduced by Christodoulou and Ruffini.
\end{abstract}
\pacs{04.70.Bw, 04.40.-b, 04.20.Cv \hfill preprint number: AEI-2007-133}
\section{Introduction} \label{s:Intro}
There are several equivalent formulae to express the mass $M$ of a single Kerr-Newman black hole in vacuum. Since in this spacetime no additional matter sources are present the mass can 
\begin{enumerate}
	\item be read off from the far-field expansion of the metric (this gives the ADM mass)
	\item be written in terms of a Komar-like integral over the black hole horizon \cite{Komar59}
	\item be expressed in terms of the angular momentum $J$, the charge $Q$ and the horizon area $A$ (Christodoulou and Ruffini \cite{ChristoRuff71}).
\end{enumerate}
If one considers axisymmetric and stationary black holes surrounded by matter sources these three descriptions are no longer equivalent. Obviously, the ADM mass characterizes the total mass of the system and is therefore not a valid measure for a local mass of the black hole. But also the mass expressions (ii) and (iii) can be very different since, e.g., in contrast to the positive definite Christodoulou-Ruffini mass, the Komar-like integral can become negative \cite{AnsPetr06}.

A well known relation for degenerate Kerr-Newman black holes is given by
\beq\label{eq:aQM}
	 a^2+Q^2=M^2,
\eeq
where $a=J/M$ is the black hole's intrinsic angular momentum per unit mass and $Q$ is its electric charge. Here we define the degeneracy of a black hole by requiring that the surface gravity $\kappa$ be zero, i.e. we follow the second characterization in the recent paper by Booth and Fairhurst \cite{BoothFair07}.

Relation (\ref{eq:aQM}) is no longer true if we take for $M$ the Komar mass parameter and allow for matter surrounding the black hole. In fact, it has been demonstrated that the Komar mass parameter can vanish for a rotating, uncharged and degenerate black hole with finite angular momentum $J$ \cite{AnsPetr06, AnsPetr07}.

In this paper we show that, in contrast, relation (\ref{eq:aQM}) holds even when the degenerate black hole is surrounded by additional matter if one takes the Christodoulou-Ruffini mass definition instead, which is given through the angular momentum $J$, the electric charge $Q$ and the horizon area $A$ by \cite{ChristoRuff71}
\beq\label{eq:ChristRuffMass}
	M=\sqrt{\left(M_\mathrm{irr} + \frac{Q^2}{4M_\mathrm{irr}}\right)^2 
		+ \frac{J^2}{4M_\mathrm{irr}^2}}\;,
\eeq
where the {\em irreducible} mass $M_\mathrm{irr}$ mass reads as
\beq\label{eq:irrMass}
	M_\mathrm{irr} = \sqrt{\frac{A}{16\pi}}.
\eeq
Thus, for the Christodoulou-Ruffini mass definition we prove that the second degeneracy characterization in \cite{BoothFair07} implies the first one in that paper. In fact, the validity of relation (\ref{eq:aQM}) distinguishes the Christodoulou-Ruffini mass parameter among other known quasi-local mass definitions, as e.g. the Hawking mass. A similar statement applies to the particular form of the angular momentum, being composed of Komar and specific electromagnetic field contributions, see section 4.

The Christodoulou-Ruffini mass conception plays a fundamental role in the isolated and dynamical horizon formalism. (See \cite{Ashtekar2005} for an overview.) Moreover, this mass parameter is being used widely in the field of dynamical calculations of spacetimes containing black holes, e.g. for describing black holes in the centre of surrounding accretion disks. (See \cite{LivRewFont03} and references therein.) In particular, a black hole is assumed to be close to extreme (i.e. degenerate) if (\ref{eq:aQM}) is nearly satisfied. However, for a dynamical spacetime or a stationary spacetime with a considerable amount of additional matter the validity of this assumption has not been demonstrated. 

In this paper we provide a justification for this assumption by showing the validity of (\ref{eq:aQM}) in the case of a single rotating, charged and degenerate black hole surrounded by additional matter in axisymmetry and stationarity. Note that, although the charge $Q$ can be neglected for astrophysically relevant situations, it is important to derive the relation (\ref{eq:aQM}) within the full Einstein-Maxwell theory since the surrounding matter may well carry a significant electromagnetic field.

The paper is organized as follows. In section \ref{s:EM-Equations}, we present the line element and the Einstein-Maxwell equations in axisymmetry and stationarity. Metric and electromagnetic expressions which stay regular in the vicinity of the black hole horizon are introduced in section \ref{s:Regular}. In section \ref{s:JQA}, we give integral formulae for the quantities needed in (\ref{eq:aQM}). The horizon boundary values in the degenerate limit, being essential for proving (\ref{eq:aQM}), are derived in section \ref{s:degenerate}. We conclude with a theorem summarizing our results and a discussion of physical inequalities for non-degenerate black holes closely related to (\ref{eq:aQM}).

We use units in which the gravitational constant and the speed of light are unity. Note that derivatives with respect to a radial coordinate, $r$ or $R$, as well as with respect to the angular coordinate $\theta$ are indicated by corresponding subscripts.
\section{Stationary and axisymmetric vacuum Einstein-Maxwell equations} \label{s:EM-Equations}
For a stationary and axisymmetric spacetime containing a black hole and 
surrounding matter, we can write the metric tensor in the electrovacuum region in terms of 
spherical coordinates $(r,\theta,\varphi,t)$ as follows:\footnote{This formulation is an adaptation of the line element used in \cite{Bardeen73}, with spherical coordinates introduced through $\rho=r\sin\theta, z=r\cos\theta$.}
\[ 
	ds^2 = e^{2\mu}\left(d r^2 + r^2 d\theta^2 \right) + 
		r^2 B^2 e^{-2\nu}\sin^2\theta \left(d\varphi - \omega\, dt\right)^2-e^{2\nu}\,dt^2.
\]
We decompose the potential $\nu$ in terms of a function $u$:
\beq\label{eq:def_u}
	\nu = u + \ln B.
\eeq
In this way  we obtain a quadruple $(\mu,B,u,\omega)$ of metric coefficients (depending on $r$ and $\theta$) which remain finite in the vicinity of the black hole, also when a central non-degenerate black hole is present\footnote{In section \ref{s:Regular} we shall introduce appropriate coordinates and functions of these metric coefficients which stay regular even in the degenerate limit.}.

Note that at the rotation axis ($\sin\theta=0$) the following regularity condition holds:
\beq\label{eq:AxisCondition}
	\mu + u = 0.
\eeq
As investigated by Carter (see \cite{Carter73}, p. 150) and Bardeen (see \cite{Bardeen73}, p. 251), the space in the immediate exterior vicinity of the black hole horizon must be electrovacuum, since any stationary matter configuration would have to resist an infinite gravitational acceleration on the horizon.
Therefore, in the vicinity of the black hole the energy momentum tensor is determined solely by the axisymmetric and stationary electromagnetic field:
\[
	4\pi T^{\mathrm (EM)}_{mn} = F_{am}F^a_{\;n}-\frac{1}{4}g_{mn}F_{ab}F^{ab}\;,
\]
where the electromagnetic field tensor can, in Lorenz gauge\footnote{Although this gauge is in most textbooks erroneously attributed to the Dutch physicist H. A. Lorentz, it was actually published by the Danish physicist L. Lorenz \cite{Lorenz1867}.}, 
be written in terms of a covector $(A_n)=(0,0,A_{\varphi},A_t)$:
\[ F_{mn} = A_{m,n}-A_{n,m}.\]
\noindent
The Einstein equations in electrovacuum are then given by:
\bea
	\nonumber\fl
		{\boldmath\nabla} \cdot \left( B {\boldmath\nabla} \nu \right) &=& 
			\frac{1}{2}\,e^{-4\nu}r^2 B^3\left({\boldmath\nabla} \omega \right)^2 \sin^2\theta
			+\frac{e^{2\nu}}{B\,r^2\sin^2\theta}\left({\boldmath\nabla} A_\varphi \right)^2 \\[2mm]
	&&\label{eq:nu}\fl\qquad\qquad\qquad\qquad
			+ Be^{-2\nu}\left({\boldmath\nabla} A_t + \omega {\boldmath\nabla} A_\varphi\right)^2 \\[2mm]
	\label{eq:om}\fl
		{\boldmath\nabla}\cdot\left(r^2 B^3 e^{-4\nu}\sin^2\theta\;{\boldmath\nabla}\omega\right) &=& 
		4\,B\,e^{-2\nu}{\boldmath\nabla}
			A_\varphi\cdot\left({\boldmath\nabla}A_t+\omega{\boldmath\nabla}A_\varphi\right)		\\[2mm]
	\label{eq:B}\fl
		{\boldmath\nabla}\cdot\left(r\sin\theta\;{\boldmath\nabla} B\right) &=& 0 \\[2mm]
	\nonumber\fl
		\mu_{rr}+\frac{1}{r}\,\mu_{r}+\frac{1}{r^2}\mu_{\theta\theta} &=& 
		\frac{1}{4}e^{-4\nu}r^2 B^2 \sin^2\theta\left({\boldmath\nabla} \omega \right)^2 \\[2mm]
	&&
	\label{eq:mu}\fl\qquad\qquad\qquad\qquad
		+ {\boldmath\nabla} \nu \cdot {\boldmath\nabla}	\left[\ln (r\sin\theta) - u\right].\\ \nonumber
\eea
\noindent
The vacuum Maxwell-equations can be written as: \footnote{Note that there are typos in the corresponding formulae in \cite{Carter73, MacdonaldThorne82}.}
\bea \label{eq:Gauss_Vac}{\boldmath\nabla} \cdot \left[ Be^{-2\nu}\left({\boldmath\nabla}A_t+\omega{\boldmath\nabla}A_\varphi\right)\right]
		&=& 0 \\
	\label{eq:Amp_Vac}{\boldmath\nabla} \cdot \left(\frac{e^{2\nu}}{B\,r^2\sin^2\theta}\,{\boldmath\nabla}A_\varphi\right)
		&=& B e^{-2\nu}{\boldmath\nabla}\omega\cdot\left({\boldmath\nabla}A_t+\omega{\boldmath\nabla}A_\varphi\right).
\eea
Here we have made use of the following identities:
\bea
	\nonumber
	 {\boldmath\nabla}\phi\cdot{\boldmath\nabla}\psi &=&\phi_{r}\psi_{r}+\frac{1}{r^2}\phi_{\theta}\psi_{\theta}\\
	\nonumber
	 {\boldmath\nabla}\cdot\left(\phi{\boldmath\nabla}\psi\right) &=&
		 \phi\left[\psi_{rr}+\frac{2}{r}\,\psi_{r}+\frac{1}{r^2}\left(\psi_{\theta\theta} + \psi_{\theta}\cot\theta\right)\right]
		+{\boldmath\nabla}\phi\cdot{\boldmath\nabla}\psi.
\eea
The ${\boldmath\nabla}$-operator has its usual meaning in three-dimensional flat space, and the functions $\phi$ and $\psi$ depend on $r$ and $\theta$ only. 
\section{Regular metric coefficients in the vicinity of the horizon} \label{s:Regular}

We follow the treatment of the general properties of stationary and axisymmetric event horizons discussed in \cite{Hawking73}, \cite{Carter73} and in particular \cite{Bardeen73}. Let, in our coordinates, the central black hole's horizon be described by a constant radius $r=r_{\mathrm h}$ and introduce a new radial coordinate $R$ through
\beq\label{eq:R}
R = \frac{1}{2}(r + \frac{r_\mathrm{h}^2}{r}).
\eeq
Then, as stated by Bardeen \cite{Bardeen73}, pp. 251, 252, the following functions of the above potentials are positive and regular with respect to $R$ and $\cos\theta$ in the vicinity of the black hole even when the degenerate limit is encountered:
\footnote{The quantities introduced here are closely related to Bardeen's expressions: $h=2r_\mathrm{h}, \lambda=2R, B_R=\hat{B}/2$}
\bea
	\label{eq:hat_mu} \hat{\mu}        &=& r^2 e^{2 \mu} \\
	\label{eq:hat_u}  \hat{u}          &=& r^2 e^{-2 u} \\
	\label{eq:hat_B}  \hat{B}          &=& \frac{r}{\sqrt{R^2-r_\mathrm{h}^2}} B.
\eea
Note that, for a single Kerr-Newman black hole in electrovacuum, the coordinate $R$ is closely related to the radial Boyer-Lindquist coordinate, $r_\mathrm{BL} = 2 R + M$. 

As for the Boyer-Lindquist form of the Kerr-Newman metric, also in the general case with surrounding matter the coordinate $R$ penetrates the horizon, that is, spatial points with coordinate values $R<r_\mathrm{h}$ are inside the horizon. Note that, in contrast, for any value $r > 0$ we obtain $R\geq r_\mathrm{h}$, i.e. the coordinate $r$ is not horizon penetrating.\footnote{In terms of the coordinates $(r,\theta,\varphi,t)$, the metric possesses an inversion symmetry with respect to the horizon, that is, the coordinate locations $(r,\theta,\varphi,t)$ and $(r^2_\mathrm{h}/r,\theta,\varphi,t)$ are physically completely equivalent.}

The remaining metric function $\omega$ as well as the electromagnetic potentials $A_\varphi$ and $A_t$ are regular in the vicinity of the horizon in terms of $R$ and $\cos\theta$. In particular, the following boundary conditions hold at $R=r_{\mathrm h}$:
\bea
	\label{BC:om}\omega &=& \mbox{constant}=\omega_{\mathrm h}\,,\;\mbox{angular velocity of the horizon} \\ 
	\label{BC:Phi}A_t+\omega_{\mathrm h} A_\varphi&=&\mbox{constant} = \Phi_{\mathrm h} \\
	\label{BC:kappa}r_{\mathrm h}\,\hat{B}\left(\hat{\mu}\hat{u}\right)^{-1/2} &=& \mbox{constant}=\kappa.
\eea
It is worthwhile stressing the constant horizon values of the functions $\omega$ and $\Phi$, where the comoving electric potential $\Phi$ is given through
\beq\label{eq:Phi}
	\Phi = A_t + \omega A_\varphi.
\eeq
The constant $\kappa$ is called the surface gravity of the horizon. We characterize a degenerate non-vanishing black hole through $\kappa=0$ with finite horizon area $A$.

For later use we introduce additional regular horizon potentials:
\bea
	\label{eq:hat_om}\hat{\omega}     &=& \frac{\omega-\omega_\mathrm{h}}{R-r_\mathrm{h}} \\
	\label{eq:hat_Phi}\hat{\Phi}      &=& \frac{\Phi-\Phi_\mathrm{h}}{R-r_\mathrm{h}}.
\eea
In \ref{a:Regular} the collection of regular potentials corresponding to the Kerr-Newman solution is given.

\section{Angular momentum, charge and horizon area of the black hole} \label{s:JQA}

In \cite{Carter73}, eqn. (9.22) (see also eqns. (9.8), (9.20) and (9.21) of that paper), the total angular momentum of the stationary axisymmetric system is given in terms of matter, electromagnetic field and black hole contributions,
\beq\label{Jtotal}
	J_\mathrm{total} = J_M + J_F + J_H
\eeq
where
\bea\label{JM}
	J_M &=& \int T^a_{Mb}m^b d\Sigma_a \\
    \label{JF}
	J_F &=& \int m^c A_c j^a d\Sigma_a + \frac{1}{4\pi} \oint_{\cal H}m^c A_c F^{ab}dS_{ab}\\
    \label{JH}
	J_H &=& \frac{1}{8\pi} \oint_{\cal H} m^{a;b} dS_{ab}.
\eea
Here, the vector $m^a$ denotes the Killing vector with respect to axisymmetry, $T^a_{Mb}$ the matter contribution in the energy momentum tensor and $j^a$ the electromagnetic current.

The volume integrals appearing in (\ref{JM}) and (\ref{JF}) are to be taken over a spacelike hypersurface $\Sigma$ expanding from the horizon $\cal H$ out to infinity.

Clearly, with the definition (\ref{Jtotal}) we cannot measure the local angular momentum of the central black hole since (\ref{Jtotal}) contains also the contribution of the surrounding matter. However, the value of $J_H$ alone can also not be used for deriving the relation (\ref{eq:aQM}) in question. For the Kerr-Newman solution this term is not equivalent to the total angular momentum, and thus (\ref{eq:aQM}) does not hold if one takes $J_H$.

The correct measure $J$ of the local angular momentum, for which (\ref{eq:aQM}) turns out to be true even in the presence of matter, is given through the combination (\ref{Jtotal}), neglecting the volume integrals appearing in (\ref{JM}) and (\ref{JF}). Due to the Einstein-Maxwell vacuum equations, this is equivalent to saying that, for the local angular momentum, the surface integrals in (\ref{JF}) and (\ref{JH}) are to be taken over some spacelike simply connected hypersurface encompassing the horizon and containing no matter sources.

In our formulation, this measure $J$ of the local angular momentum leads to the horizon surface integral:
\bea\nonumber
	J&=&-\frac{1}{16\pi}\oint_{\cal H} \left(e^{-4\nu}r^2 B^3\sin^2\theta \right) {\boldmath\nabla}\omega\cdot d\vec{f} 
		\\ && \nonumber
		+\frac{1}{4\pi}\oint_{\cal H} 
		\left[ Be^{-2\nu}A_\varphi\left({\boldmath\nabla}A_t+\omega{\boldmath\nabla A_\varphi}\right)\right]
			\cdot d\vec{f}\\
		&=& \label{eq:J}\left.\int_0^\pi\frac{\hat{u}}{2\hat{B}}\left[-\frac{1}{4}\,\hat{\omega}\,\hat{u}\sin^2\theta
			+ A_\varphi(\hat{\Phi}-A_\varphi\;\hat{\omega})\right] \sin\theta d\theta \right|_{R=r_{\mathrm h}}.
\eea
Note that in the presence of an electromagnetic field this local angular momentum parameter differs from the formula (15) in \cite{BoothFair07} which only involves metric but no electromagnetic quantities \footnote{For the spacetimes considered here, formula (15) in \cite{BoothFair07} is equivalent to (\ref{JH}).}.

The local charge $Q$ of the black hole can be defined unambigiously by the surface integral (9.25) of \cite{Carter73}, which reads in our formulation as follows:
\bea
	Q &=& -\frac{1}{4\pi}\oint_{\cal{H}} F^{ab}dS_{ab}=
		-\frac{1}{4\pi}\oint_{\cal{H}} Be^{-2\nu}\left({\boldmath\nabla}A_t+\omega{\boldmath\nabla}A_\varphi\right)\nonumber
		\cdot d\vec{f}\\
		\label{eq:Q}
		&=& \left.-\int_0^\pi\frac{\hat{u}}{2\hat{B}}(\hat{\Phi}-A_\varphi\;\hat{\omega})\sin\theta d\theta\right|_{R=r_{\mathrm h}}.
\eea
Finally, the horizon area is given through the expression:
\beq\label{eq:A}
	A = 2\left.\pi\int_0^\pi \sqrt{\hat{\mu}\hat{u}}\sin\theta d\theta\right|_{R=r_{\mathrm h}}.
\eeq

\section{The degenerate limit} \label{s:degenerate}

As mentioned above, we describe a degenerate black hole through the vanishing of the surface gravity $\kappa$ which implies, by virtue of (\ref{BC:kappa}), that $r_{\mathrm h}=0$ in this limit. Remarkably, it is then possible to work out explicitly the horizon boundary values of all functions $\hat{u}, \hat{\mu}, \hat{B}, \hat{\omega}, \hat{\Phi}$ and $A_\varphi$ because, in the Einstein-Maxwell-equations evaluated at the horizon $R=r_{\mathrm h}=0$, all terms containing $R$-derivatives disappear.

We will use this information to determine the integrals presented in the previous section and ultimately to verify the validity of (\ref{eq:aQM}).

If we express the Einstein-Maxwell-equations (\ref{eq:nu} -- \ref{eq:Amp_Vac}) in terms of the coordinates $R,\theta$ and the potentials $\hat{u}, \hat{\mu}, \hat{B}, \hat{\omega}, \hat{\Phi}$ and $A_\varphi$, we find:
\begin{enumerate}

\item Equation (\ref{eq:nu}):\\ (For simplicity, we write the equation for $\tilde{u}=\frac{1}{2}\ln\hat{u}$ rather than for $\hat{u}$.)
\bea
\fl\nonumber\nopagebreak
\lefteqn{(R^2-r_\mathrm{h}^2) [\tilde{u}_{RR}+\tilde{u}_R(\ln\hat{B})_R] + 2 R \tilde{u}_R + \tilde{u}_{\theta\theta} + \tilde{u}_\theta[\cot\theta + (\ln\hat{B})_\theta] +(\ln\hat{B})_\theta \cot\theta}&&\\ \nopagebreak
\fl\nonumber&&\fl=\label{Eqn_for_ud} \quad
1-\frac{e^{4\tilde{u}}}{2\hat{B}^2}\sin^2\theta\left[\omega^2_R+\frac{R-r_\mathrm{h}}{R+r_\mathrm{h}}\hat{\omega}^2_\theta\right]
-\frac{e^{-2\tilde{u}}}{\sin^2\theta}\left[(R^2-r_\mathrm{h}^2)A_{\varphi,R}^2+A_{\varphi,\theta}^2\right] \\ && \fl\qquad\nopagebreak
-\frac{e^{2\tilde{u}}}{\hat{B}^2}\left[(\Phi_R-A_\varphi\omega_R)^2
	+\frac{R-r_\mathrm{h}}{R+r_\mathrm{h}}(\hat{\Phi}_\theta-A_\varphi\hat{\omega}_\theta)^2\right]
\eea

\item Equation (\ref{eq:om}):
\bea
\fl\nonumber\nopagebreak
\lefteqn{(R+r_\mathrm{h})\left(\omega_{RR}+\omega_R\left[4\tilde{u}_R-(\ln\hat{B})_R\right]\right)
	+\hat{\omega}_{\theta\theta} 
	+ \hat{\omega}_\theta\left(3\cot\theta+4\tilde{u}_\theta-(\ln\hat{B})_\theta\right)
}&&\\ \label{Eqn_for_omd}
\fl&&=\frac{4 e^{-2\tilde{u}}}{\sin^2\theta}\left[(R+r_\mathrm{h})A_{\varphi,R}(\Phi_R-A_\varphi\omega_R)
	+ A_{\varphi,\theta}(\hat{\Phi}_\theta-A_\varphi\hat{\omega}_\theta)\right]
\eea

\item Equation (\ref{eq:B}):
\beq \label{Eqn_for_Bd}
	(R^2-r_\mathrm{h}^2) \hat{B}_{RR} +3 R \hat{B}_R + \hat{B}_{\theta\theta} + 2\hat{B}_{\theta}\cot\theta = 0\\
\eeq

\item Equation (\ref{eq:mu}):\\
 (For simplicity, we write the equation for $\tilde{\mu}=\frac{1}{2}\ln\hat{\mu}$ rather than for $\hat{\mu}$.)
\bea
\nonumber\fl
\lefteqn{(R^2-r_\mathrm{h}^2) \tilde{\mu}_{RR} + R \tilde{\mu}_R + \tilde{\mu}_{\theta\theta}}&&\\
\nonumber\fl&&\fl=
\frac{e^{4\tilde{u}}}{4\hat{B}^2}\sin^2\theta\left[\omega^2_R+\frac{R-r_\mathrm{h}}{R+r_\mathrm{h}}\hat{\omega}^2_\theta\right]
-(R^2-r_\mathrm{h}^2)\tilde{u}_R\left[\tilde{u}_R-(\ln\hat{B})_R\right] \\ &&\fl \qquad\label{Eqn_for_mud}
+R\tilde{u}_R- \left[\tilde{u}_\theta+\cot\theta\right]\left[\tilde{u}_\theta-(\ln\hat{B})_\theta\right]
\eea

\item Equation (\ref{eq:Gauss_Vac}):
\bea
\fl\nonumber
\lefteqn{(R+r_\mathrm{h})\left[\left(\Phi_{RR}-A_\varphi\omega_{RR}\right)+\left(\Phi_{R}
	-A_\varphi\omega_{R}\right)\left(2\tilde{u}_R-(\ln\hat{B})_R\right)
-A_{\varphi,R}\omega_{R}\right]
}&&\\ \label{Eqn_for_Phid}
\fl&&+\left(\hat{\Phi}_{\theta\theta}-A_\varphi\hat{\omega}_{\theta\theta}\right)
+\left(\hat{\Phi}_{\theta}-A_\varphi\hat{\omega}_{\theta}\right)\left(2\tilde{u}_\theta
	-(\ln\hat{B})_\theta+\cot\theta\right)-A_{\varphi,\theta}\hat{\omega}_{\theta}\quad=0
\eea

\item Equation (\ref{eq:Amp_Vac}):
\bea
\nonumber
\lefteqn{(R^2-r_\mathrm{h}^2) \left[A_{\varphi,RR}+A_{\varphi,R}\left(-2\tilde{u}_R+(\ln\hat{B})_R\right)\right] + 2 R A_{\varphi,R}} &&\\ 
\nonumber &&
 +A_{\varphi,\theta\theta} + A_{\varphi,\theta}\left[-2\tilde{u}_\theta 
	+ (\ln\hat{B})_\theta-\cot\theta\right] \\
&&=\label{Eqn_for_Aphi}
\frac{e^{4\tilde{u}}}{\hat{B}^2}\sin^2\theta\left[\omega_R(\Phi_R-A_\varphi\omega_R)
	+\frac{R-r_\mathrm{h}}{R+r_\mathrm{h}}\hat{\omega}_\theta(\hat{\Phi}_\theta-A_\varphi\hat{\omega}_\theta)\right].
\eea

\end{enumerate}
As already mentioned, it becomes apparent that at the horizon $R=r_\mathrm{h}$ in the degenerate limit $r_\mathrm{h}=0$ all terms containing $R$-derivatives disappear\footnote{In order to see this, in (\ref{Eqn_for_ud}-\ref{Eqn_for_Aphi}) all occurrences of the potentials $\omega$ and $\Phi$ need to be replaced by their counterparts $\hat{\omega}$ and $\hat{\Phi}$.}. This means that we can explicitly calculate the regular horizon values of the metric potentials.

First, (\ref{Eqn_for_Bd}) tells us that 
\[
	\hat{B}_{\theta\theta} + 2\hat{B}_{\theta}\cot\theta = 0\\
\]
which leads to a constant horizon value of $\hat{B}$, being the only regular solution of this equation. Note that in the presence of surrounding matter this value $\hat{B}$ is in general different from the Kerr-Newman value ($\hat{B}_\mathrm{Kerr-Newman}=2$, see \ref{a:Regular}).

Using this result, we can write (\ref{Eqn_for_Phid}) for $R=r_\mathrm{h}=0$:
\[
(\hat{\Phi}_{\theta\theta}-A_\varphi\hat{\omega}_{\theta\theta}-A_{\varphi,\theta}\hat{\omega}_{\theta})
+\left(\hat{\Phi}_{\theta}-A_\varphi\hat{\omega}_{\theta}\right)\left(2\tilde{u}_\theta
	+\cot\theta\right)=0.
\]
The only regular solution of this equation reads:
\[\hat{\Phi}_{\theta}=A_\varphi\hat{\omega}_{\theta}.\]
If we insert this result in (\ref{Eqn_for_omd}), together with the constant horizon value of $\hat{B}$, we obtain
\[
	\hat{\omega}_{\theta\theta} 
	+ \hat{\omega}_\theta\left(3\cot\theta+4\tilde{u}_\theta\right)=0.
\]	
It therefore follows that, apart from $\hat{B}$, also $\hat{\omega}$ and $\hat{\Phi}$ are constant at the horizon in the degenerate limit. Let us call these constants $\hat{B}_\mathrm{h},\hat{\omega}_\mathrm{h}$ and $\hat{\Phi}_\mathrm{h}$ respectively, with $\hat{B}_\mathrm{h}>0$ in accordance with the regularity properties discussed in section \ref{s:Regular}.
With this result consider equations (\ref{Eqn_for_ud}) and (\ref{Eqn_for_Aphi}) for $R=r_\mathrm{h}=0$:
\bea
\nopagebreak\nonumber\tilde{u}_{\theta\theta} + \tilde{u}_\theta\cot\theta &=& 
1-\frac{e^{4\tilde{u}}\hat{\omega}_\mathrm{h}^2}{2\hat{B}_\mathrm{h}^2}\sin^2\theta \nopagebreak\\&&\nopagebreak\label{Eqn_for_ud_2}\nopagebreak
-\frac{e^{-2\tilde{u}}A_{\varphi,\theta}^2 }{\sin^2\theta}
-\frac{e^{2\tilde{u}}}{\hat{B}_\mathrm{h}^2}(\hat{\Phi}_\mathrm{h}-A_\varphi\hat{\omega}_\mathrm{h})^2\\
\nopagebreak\label{Eqn_for_Aphi_2} A_{\varphi,\theta\theta} - A_{\varphi,\theta}(2\tilde{u}_\theta +\cot\theta) &=&
\frac{e^{4\tilde{u}}}{\hat{B}_\mathrm{h}^2}\hat{\omega}_\mathrm{h}(\hat{\Phi}_\mathrm{h}-A_\varphi\hat{\omega}_\mathrm{h})\sin^2\theta.
\eea	
Here we have used that according to  (\ref{eq:hat_om}, \ref{eq:hat_Phi}):
\[
	\begin{array}{lclclcl}
		\omega_R(R=r_\mathrm{h},\theta) &=& \hat{\omega}(R=r_\mathrm{h},\theta)
			&\to&\hat{\omega}_\mathrm{h}&\mbox{as} & r_\mathrm{h}\to 0\\ 
  		\Phi_R(R=r_\mathrm{h},\theta)   &=& \hat{\Phi}(R=r_\mathrm{h},\theta)
		&\to&\hat{\Phi}_\mathrm{h}&\mbox{as} & r_\mathrm{h}\to 0.
	\end{array}
\]

In analogy to the solutions for the Kerr-Newman case (see \ref{a:Regular}) we expect that (\ref{Eqn_for_ud_2}, \ref{Eqn_for_Aphi_2}) have a special solution of the form:
\bea
	 \nonumber\hat{u}_\mathrm{h}     &=& \frac{c_1}{1 + \alpha^2\cos^2\theta} \\
     \nonumber(A_\varphi)_\mathrm{h} &=& \frac{c_2\sin^2\theta}{1 + \alpha^2\cos^2\theta}.
\eea
Insertion into (\ref{Eqn_for_ud_2}, \ref{Eqn_for_Aphi_2}) confirms this supposition, and fixes the constants $c_1, c_2$ and $\alpha$ such that we have (for $\hat{\omega}_\mathrm{h}\neq 0$)
\bea
	\label{uh_2}\hat{u}_\mathrm{h}   &=& \frac{2\,\hat{B}_\mathrm{h}\alpha}{\hat{\omega}_\mathrm{h}(1+\alpha^2 \cos ^2\theta)}\\
	\label{Aphih_2}(A_\varphi)_\mathrm{h} &=& -\frac{2\,\hat{\Phi}_\mathrm{h}\,\alpha^2\sin^2\theta}
		{\hat{\omega}_\mathrm{h}(1-\alpha^2)(1+\alpha^2 \cos ^2\theta)}\;
\eea
with the parameter $\alpha$ determined by
\beq\label{Relation_alpha_2}
	\hat{B}_\mathrm{h}\hat{\omega}_\mathrm{h}(1-\alpha^2)^3
	= 2\alpha(1+\alpha^2)^2\hat{\Phi}_\mathrm{h}^2.
\eeq
Although (\ref{Relation_alpha_2}) is nonlinear in $\alpha$, it defines $\alpha\in[-1,1]$ uniquely through $\hat{B}_\mathrm{h},\hat{\omega}_\mathrm{h}$ and $\hat{\Phi}_\mathrm{h}$ because
\[
\frac{\alpha(1+\alpha^2)^2}{(1-\alpha^2)^3}
\]
is monotonically increasing with respect to $\alpha$ in the range $(-1,1)$. Note that $\alpha^2 =1$ signifies the uncharged case and $\alpha = 0$ the non-rotating case. The signs of $\alpha$ and $\hat{\omega}_\mathrm{h}$ always coincide which ensures the positive definiteness of $\hat{u}_\mathrm{h}$, see section \ref{s:Regular}.

It seems rather involved to prove that (\ref{uh_2}, \ref{Aphih_2}) is the unique regular solution of the nonlinear, coupled
system of Einstein-Maxwell equations (\ref{Eqn_for_ud_2}, \ref{Eqn_for_Aphi_2}).
In \ref{NonLinSystem} we rewrite equations (\ref{Eqn_for_ud_2}, \ref{Eqn_for_Aphi_2}) such that the physical constants $\hat{B}_\mathrm{h},\hat{\omega}_\mathrm{h}$ and $\hat{\Phi}_\mathrm{h}$ are eliminated, thus obtaining a particularly clear mathematical structure of this system. In the framework of this modified system we present some steps towards its general solution, although a complete study is beyond the scope of this paper.
It is, however, possible to prove that (\ref{uh_2}, \ref{Aphih_2}) is an isolated solution, i.e. that there exist no neighbouring solutions to (\ref{Eqn_for_ud_2}, \ref{Eqn_for_Aphi_2}) fulfilling the geometrical and physical conditions of regularity, equatorial symmetry, and axis regularity. (See \ref{Annex_to_deg} for details.) As will be discussed in section \ref{s:Dicussion}, we may use this result to prove relation (\ref{eq:aQM}) for any axially and equatorially symmetric stationary configuration, consisting  of a degenerate black hole and surrounding matter, which allows for a continuous parameter transition to a degenerate Kerr-Newman black hole in vacuum.

Consider now (\ref{Eqn_for_mud}) for $R=r_\mathrm{h}=0$
\[
\tilde{\mu}_{\theta\theta}=
\frac{e^{4\tilde{u}}\hat{\omega}_\mathrm{h}^2}{4\hat{B}_\mathrm{h}^2}\sin^2\theta
- \tilde{u}_\theta\left(\tilde{u}_\theta+\cot\theta\right).
\]
From the solution (\ref{uh_2}) we can determine $\tilde{\mu}$ through integration with respect to $\theta$. The corresponding constants of integration can be determined from (i) the requirement of an axially and equatorially symmetric solution (that is $\tilde{\mu}_\theta=0$ for $\theta=0$ and $\theta=\pi/2$) and (ii) from the axis regularity condition (\ref{eq:AxisCondition}). It thus follows that
\beq\label{muh_2}
	\hat{\mu}_\mathrm{h} = \frac{2\hat{B}_\mathrm{h}\alpha(1+\alpha^2 \cos^2\theta)}{\hat{\omega}_\mathrm{h}(1+\alpha^2)^2}.
\eeq
This expression is positive definite, which is again in accordance with the regularity properties discussed in section \ref{s:Regular}.

With the explicit knowledge of the horizon values of the metric and electromagnetic functions in the degenerate limit we can now express angular momentum $J$, charge $Q$ and horizon area $A$ by virtue of the formulae (\ref{eq:J}, \ref{eq:Q}, \ref{eq:A}):
\bea
	\nonumber J &=& -\frac{2 \alpha^2}{(1+\alpha^2)^2}\,\frac{\hat{B}_\mathrm{h}}{\hat{\omega}_\mathrm{h}}\\
	\nonumber Q &=& -\frac{2 \alpha}{1-\alpha^2}\,\frac{\hat{\Phi}_\mathrm{h}}{\hat{\omega}_\mathrm{h}}\\
	\nonumber A &=& \frac{8\pi \alpha}{1+\alpha^2}\,\frac{\hat{B}_\mathrm{h}}{\hat{\omega}_\mathrm{h}}.
\eea
As a consequence we find
\bea
	\nonumber\frac{4\pi J}{A}   &=& -\frac{\alpha}{1+\alpha^2}\\
	\nonumber\frac{4\pi Q^2}{A} &=& \frac{1-\alpha^2}{1+\alpha^2}\\
	\nonumber\frac{4\pi M^2}{A} &=& \frac{1}{1+\alpha^2}
\eea
where for the latter formula we have written the Christodoulou-Ruffini mass $M$ in accordance with (\ref{eq:ChristRuffMass}).
These expressions imply (\ref{eq:aQM}) with $a=J/M$.

Note that (\ref{eq:aQM}) is equivalent to the relation
\beq\label{JQA_Rot_Rate}
	p_J^2 + p_Q^2 = 1\;,
\eeq
where the parameters $p_J$ and $p_Q$ are given by
\bea
	\label{p_J} p_J &=& \frac{8\pi J}{A} \\
	\label{p_Q} p_Q &=& \frac{4\pi Q^2}{A} \geq 0.
\eea
We shall discuss this further in the next section.

About the above results the following remarks are
appropriate: In the degenerate case we originally had 4 integration
constants $\hat{B}_\mathrm{h}, \hat{\omega}_\mathrm{h}, \hat{\Phi}_\mathrm{h}$ and $\alpha$, which, according to (\ref{Relation_alpha_2}), reduce to 3 constants, e.g. $\hat{B}_\mathrm{h}, \hat{\omega}_\mathrm{h}/\hat{B}_\mathrm{h}$ and $\alpha$.
However, $\hat{B}_\mathrm{h}$ appears in the decisive differential equations (\ref{Eqn_for_ud_2}, \ref{Eqn_for_Aphi_2}) only
in the combination $\hat{\omega}_\mathrm{h}/\hat{B}_\mathrm{h}$. Since $\alpha$ is dimensionless, and
$\hat{\omega}_\mathrm{h}/\hat{B}_\mathrm{h}$ has dimension $(\mathrm{length})^{-2}$, it is clear that $J, Q^2$ and $A$
have to be proportional to $\hat{B}_\mathrm{h}/\hat{\omega}_\mathrm{h}$, and that in the dimensionless
quantities $p_J$ and $p_Q$ this dependence cancels. The fact that these quantities can be combined to give the universal relation (\ref{JQA_Rot_Rate}) is, however, highly non-trivial.

\section{Discussion} \label{s:Dicussion}
In this section we study at first physical consequences of the mathematical derivation presented in the previous section. In particular we are able to prove a theorem about the validity of relation (\ref{eq:aQM}) for degenerate black holes with surrounding matter. In a second part we discuss closely related specific physical inequalities for non-degenerate black holes.

\begin{enumerate}
	\item 
	Let us consider a configuration, consisting of a degenerate central black hole with surrounding matter, which is sufficiently close to the Kerr-Newman solution. If we assume a physically reasonable type of matter, i.e. possessing positive baryonic mass density, we may characterize this configuration by saying that the positive parameter $P$,
	\beq\label{eq:P}
		P = \frac{M_\mathrm{B}}{M_\mathrm{irr}} \;,
	\eeq
	is sufficiently small; here $M_\mathrm{B}$ denotes the total baryonic mass of the surrounding matter. For such a system all metric and electromagnetic potentials differ only slightly from the Kerr-Newman expressions. As a consequence, from (\ref{Eqn_for_ud_2}, \ref{Eqn_for_Aphi_2}) again (\ref{uh_2}, \ref{Aphih_2}) emerges since this solution is isolated. Therefore we may conclude that the relations (\ref{eq:aQM}) and (\ref{JQA_Rot_Rate}) also hold for the considered configuration in the vicinity of the Kerr-Newman black hole.
	
	Let us now consider such a system with prescribed $P=P_0>0$ permitting the identification of a continuous sequence of configurations parameterized by $P$, with the values $P=P_0$ and $P=0$ denoting start and end points of the sequence. This is to say that the baryonic mass of the original system can be decreased parametrically such that eventually a degenerate Kerr-Newman solution is encountered. Along this transition all black hole parameters are assumed to depend continuously on $P$. If for a system such a continuous parameter transition can be found, then we may conclude the validity of the relations (\ref{eq:aQM}) and (\ref{JQA_Rot_Rate}). The reason for this is again the fact that (\ref{uh_2}, \ref{Aphih_2}) is isolated. Let us assume, in contrast, that the continuous function
	\beq\label{eq:S}
		S = p_J^2 + p_Q^2\;,
	\eeq
	depending on $P$, is not identically 1. Then there is a point $P_1>0$ and a positive real number $\epsilon$ such that 
	\begin{enumerate}	
		\item $S(P)=1$ for $P\in(P_1-\epsilon,P_1)$ with the solutions (\ref{uh_2}, \ref{Aphih_2}) of (\ref{Eqn_for_ud_2}, \ref{Eqn_for_Aphi_2}) ,
		\item $S(P)\neq 1$ for $P\in(P_1,P_1+\epsilon)$.
	\end{enumerate}
	For $P\in(P_1,P_1+\epsilon)$, the solution of (\ref{Eqn_for_ud_2}, \ref{Eqn_for_Aphi_2}) must depart continuously from (\ref{uh_2}, \ref{Aphih_2}) in order to realize the required continuous parametriziation through $P$. However, (\ref{uh_2}, \ref{Aphih_2}) is isolated, and therefore any other solution cannot depart continuously from it. 

	These considerations are also valid if the sequence in question is parametrized through a different quantity $\tilde{P}$. The only requirement needed is that the horizon values of the metric and electromagnetic potentials, $\hat{u}, \hat{\mu}, \hat{B}, \hat{\omega}, \hat{\Phi}$ and $A_\varphi$, depend continuously on the parameter being chosen.\\
	
Thus we have proved the following \\
	
	{\bf Theorem:}\nopagebreak \\[2mm]\nopagebreak
		{\em
			Consider an axially and equatorially symmetric, stationary configuration consisting of a degenerate central black hole with surrounding matter. Assume that a parametric sequence of such configurations can be identified that describes a continuous transition of the horizon values of the metric and electromagnetic potentials from the given system to those of a single Kerr-Newman black hole in electrovacuum. Then the relations (\ref{eq:aQM}) and (\ref{JQA_Rot_Rate}) hold along the entire sequence, including the original configuration.\\[1mm]
		}
	\begin{center}
		\begin{figure}[h]
			\hspace*{0.7cm}\includegraphics[scale=0.82]{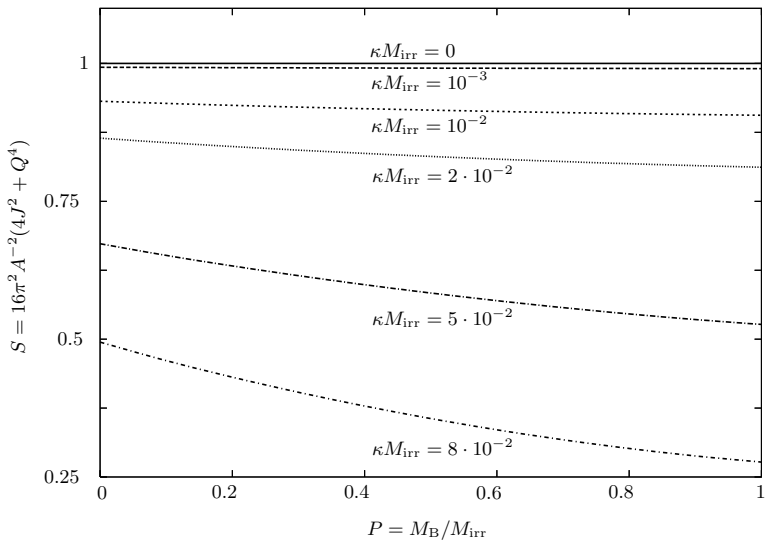}
			\caption{
			Sequences of configurations consisting of a charged central black hole and a surrounding 
			infinitely flattened, uniformly rotating ring of electrically neutral dust matter. All sequences describe a transition to the Kerr-Newman black hole ($P=0$; $M_\mathrm{B}$ is the ring's baryonic mass). For rigid rotation of the ring this transition can only be achieved if in the Kerr-Newman limit the ratio of inner to outer circumferential radius of the ring ($\varrho_\mathrm{circ} = r e^{-u}$) tends to 1, since the particles of a `test-'ring of finite width would have to follow the circular geodesics in the Kerr-Nemann geometry and hence cannot possess a unifom angular velocity. Here we have chosen
				\[ \frac{\varrho_\mathrm{circ,inner}}{\varrho_\mathrm{circ,outer}} = 1 - \frac{3}{10}\sqrt{P}.\]
			For all sequences we have prescribed the value for $p_Q=\frac{1}{2}$ (see (\ref{p_Q})) and a constant value for $\kappa M_\mathrm{irr}$. Note that we have chosen the black hole and the ring to be in co-rotation. The figure reveals that the value for $S$ (see (\ref{eq:S})) tends to 1 in the degenerate case $\kappa M_\mathrm{irr}=0$, independently of the value for $P$, i.e. the amount of surrounding matter. The numerical calculations have been carried out for $\kappa M_\mathrm{irr}\in\{10^{-3}, 10^{-2},2\cdot 10^{-2}, 5\cdot 10^{-2}, 8\cdot 10^{-2}\}$.
			}
			\label{Fig_1}
		\end{figure}
	\end{center}
	The existence of this parametric sequence does not appear to be a stringent physical restriction. For a matter configuration in a certain spatial distance from the black hole horizon, it should be possible to perform parametrically a transition (e.g. through the gradual decrease of the baryonic mass) such that the black hole quantities change continuously and eventually assume the Kerr-Newman expressions. For an illustration see \fref{Fig_1}, in which several sequences of this kind with exterior dust matter are considered. Also the configurations presented in \fref{Fig_2} have been obtained through a parametric sequence, driving the parameter $P$ from 0 (the Kerr-Newman value) to 1. Moreover, for the pure Einstein-field, numerical examples of such sequences have been studied in \cite{AnsPetr05}.

	Note that we have not required an asymptotic flatness condition. Indeed, it seems conceivable that a non-asymptotically flat configuration can in fact be deformed such that in a continuous manner the horizon values of a Kerr-Newman black hole are assumed. In this case again the theorem would be applicable. Particular examples of this kind will be the subject of a future publication.

	It is worthwhile to stress that we also assume the axisymmetric and stationary configurations being considered to be reflectionally symmetric with respect to the equatorial plane. While this can be shown to hold in Newtonian gravity it is still a conjecture in Einstein's theory. Nevertheless, the numerous solutions constructed analytically and numerically do all possess this property.

	\begin{figure}[h]
		\hspace*{0.7cm}\includegraphics[scale=0.82]{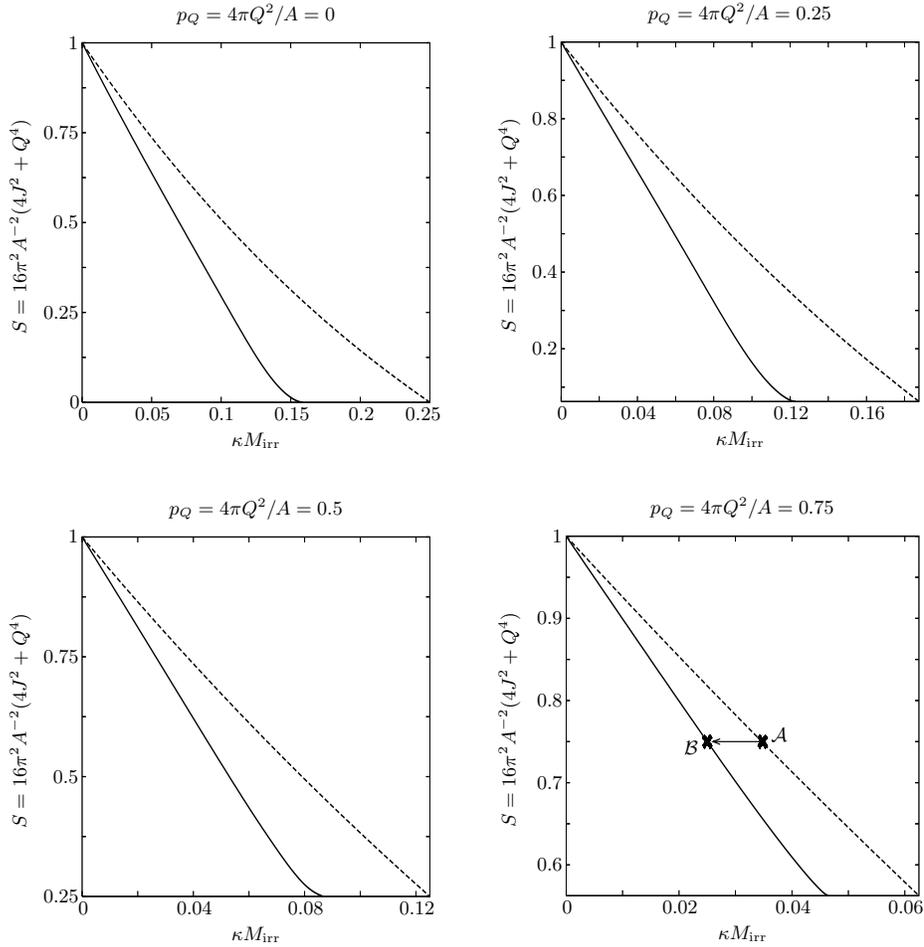}
		\caption{
		Sequences of configurations consisting of a charged central black hole and a surrounding 
		infinitely flattened, uniformly rotating ring of electrically neutral dust matter. For each of the sequences we prescribe the value 
		for $p_Q$. Moreover, the parameter defined in (\ref{eq:P}) is fixed at $P=1$, and the ratio of 
		inner to outer circumferential radius of the ring ($\varrho_\mathrm{circ} = r e^{-u}$) is taken to be 
		0.7 for all models. The sequences are considered for increasing parameter $8\pi J/A$, starting at zero, i.e. at the minimal value for the parameter $S$ (see (\ref{eq:S})). Note that we have chosen the black hole and the ring to be in co-rotation. The solid lines reveal that along these sequences the surface gravity decreases monotonically and reaches zero exactly when $S$ is equal to 1. Also shown (dashed lines) are the curves for the Kerr-Newman black hole ($P=0$). They obey the relation
		\[
			S=1 + 8\kappa^2 M_\mathrm{irr}^2 - 4\sqrt{2}\, \kappa M_\mathrm{irr} 
				\sqrt{1 + p_Q + 2\kappa^2 M_\mathrm{irr}^2}.
		\]
		The points ${\cal A}$ and $\cal{B}$ marked with crosses are referred to at the end of the discussion.
		}
		\label{Fig_2}
	\end{figure}
	\item
	From the very definition of the Christodoulou-Ruffini mass (\ref{eq:ChristRuffMass}) it follows directly that 
	\beq\label{in:aQM}
		a^2+Q^2\leq M^2
	\eeq
	holds for all (degenerate and non-degenerate) black holes with or without surrounding matter. This can most easily be seen from the identity
	\beq
		\label{a2pQ2}
		\frac{a^2+Q^2}{M^2} = 1-\left(\frac{1-S}{1+2p_Q+S}\right)^2.
	\eeq
	It is, in contrast, not known to what extent the inequality
	\beq\label{in:JQA}
		S=p_J^2 + p_Q^2 \leq 1
	\eeq
	is satisfied which is valid for the Kerr-Newman solution. 

	By means of a highly accurate numerical code based on pseudo-spectral methods we are able to investigate the validity of (\ref{in:JQA}) for systems of black holes with surrounding matter. The details of the numerical method can be found in \cite{AnsPetr05}. In particular, we have considered uniformly rotating homogeneous tori around uncharged black holes. A new version of the code treats more general fluid matter (as e.g. polytropes) and differential rotation. Apart from that, we have created a modified version that handles charged black holes surrounded by dust rings. In this manner we have investigated a large number of solutions and found that they all satisfy the relation (\ref{in:JQA}). Hence, we are led to the following \\
	
	{\bf Conjecture:}\\[2mm]
		{\em
			Consider an axially and equatorially symmetric, stationary configuration consisting of a central black hole with surrounding matter. Then the inequality (\ref{in:JQA}) holds, and the equal sign is assumed if and only if the central black hole is degenerate.\\[1mm]
		}

	Note that at the current status it is uncertain what physical conditions on the type of matter have to be required, in order for the conjecture to be applicable.\\

	The conjecture gives a partial answer to the question raised by Booth and Fairhurst at the end of section III. A of \cite{BoothFair07}.
	As an illustration we include \fref{Fig_2}  which reveals that the parameter point ($p_J, p_Q$) is indeed confined to the interior of the unit circle, reaching the circle's boundary as the degenerate limit $\kappa=0$ is encountered. We conclude that for fixed value $p_Q$ the expression (\ref{a2pQ2})
	is monotonically increasing with respect to the parameter $S$. This means that a black hole with given horizon area $A$ and charge $Q$ acquires more and more (Christodoulou-Ruffini) mass as the angular momentum is increased and assumes a final maximal mass in the degenerate limit when the equal sign in (\ref{in:JQA}) is reached.

	In this manner we encounter the familiar picture of a single Kerr-Newman black hole in vacuum. It is therefore tempting to characterize the Christodoulou-Ruffini mass as the measure of the quasi-local bare mass of the black hole, an interpretation also supported by the isolated horizon framework \cite{Ashtekar2005}. Note that the Komar mass could not be taken as a bare mass measure because this parameter incorporates the specific relativistic spacetime geometry in the black hole's vicinity, see \cite{AnsPetr07}.

	These considerations provide a justification for saying that a black hole
	is characterized through its intrinsic physical parameters angular momentum
	$J$, charge $Q$ and area $A$ and hence Christodoulou-Ruffini mass $M$. It
	therefore appears conceivable to compare different black hole matter
	configurations with the `same' central black hole described by mutual values for 
	$J, Q$ and $A$. For example, consider the two configurations ${\cal A}$ and ${\cal B}$ in \fref{Fig_2} possessing the same value $S=0.75$. If one adds a ring of matter exterior to the Kerr-Newman black hole ${\cal A}$ we may expect that the radial force to be exerted on a zero angular momentum observer (ZAMO, see \cite{Bardeen73}) decreases in the vicinity of the black hole. This is due to the gravitational attraction towards the ring. As a consequence the surface gravity $\kappa$ becomes smaller because this quantity just measures a rescaled acceleration of the ZAMO's on the black hole horizon. It thus becomes plausible that in \fref{Fig_2} the transition from ${\cal A}$ to ${\cal B}$ is described by decreasing values of the parameter $\kappa M_\mathrm{irr}$. Note that this effect becomes smaller and smaller if one approaches the degenerate black hole.
	
\end{enumerate}
In the future we shall further examine the relation (\ref{in:JQA}). In particular, we plan a detailed study of charged black hole configurations with surrounding dust and fluid matter.

\ack

It is a pleasure to thank D.~Petroff and J.~L.~Jaramillo
for numerous valuable discussions. We are in particular grateful to D.~Petroff
for his assistance in the linear analysis of equations (\ref{Eqn_for_ud_2},
\ref{Eqn_for_Aphi_2}). This work was supported by the grant SFB/Transregio 7
`Gravitational Wave Astronomy' funded by the German Research Foundation.
H.~P.~wishes to thank for the hospitality and support of the Max Planck
Institute for Gravitational Physics (Albert Einstein Institute).

\appendix
\section{Kerr-Newman expressions}\label{a:Regular}

The regular potentials discussed in section \ref{s:Regular} read for the Kerr-Newman solution as follows:
\bea
	\nonumber\hat{\mu}        &=& (M+2 R)^2+a^2 \cos ^2\theta\\
	\nonumber\hat{u}          &=& \hat{\mu}^{-1}\left(\left[a^2+(M+2 R)^2\right]^2-4a^2 \left(R^2-r_\mathrm{h}^2\right)
						   \sin ^2\theta\right)\\
	\nonumber\hat{B}          &=& 2 \\
	\nonumber\omega           &=& a \left[2 M (M+2 R)-Q^2\right] (\hat{\mu}\hat{u})^{-1}\\
	\nonumber\Phi             &=& Q(M+2R)\hat{\mu}^{-1}(1-a\,\omega\sin ^2\theta)\\
	\nonumber A_\varphi        &=& -Qa(M+2R)\hat{\mu}^{-1}\sin ^2\theta.
\eea
Here,
\[ 
	Q^2  = M^2-a^2-4r_\mathrm{h}^2.
\]
Moreover, angular momentum $J$, horizon area $A$ and surface gravity $\kappa$ read as (with the abbreviation $b=M+2 r_\mathrm{h}$),
\bea
	\nonumber J &=& M a \\
	\nonumber A &=& 4\pi (2 Mb -Q^2)\\
	\nonumber \kappa &=& \frac{2 r_\mathrm{h}}{2 Mb -Q^2}.
\eea
The horizon values turn out to be
\bea
	\nonumber\omega_\mathrm{h}         &=&\frac{a}{a^2+b^2}\\
	\nonumber\Phi_\mathrm{h}           &=&Q\frac{b}{a^2+b^2}.
\eea
The additional regular functions introduced at the end of section \ref{s:Regular} are given by:
\bea
	\nonumber\hat{\omega}     &=& \frac{4a^3(R+r_\mathrm{h})\sin^2\theta
			- \sum\limits_{i=0}^3 a_i(R-r_\mathrm{h})^i}{\hat{\mu}\hat{u}(a^2+b^2)}\\[5mm]
	\nonumber\hat{\Phi}       &=& Q\left[-\frac{M+2R}{\hat{\mu}}\left(\frac{4(M+R+r_\mathrm{h})}{a^2+b^2}
			+a\hat{\omega}\sin^2\theta\right) + \frac{2}{a^2+ b^2}\right]
\eea
with,
\bea
	\nonumber a_0&=& 4a(b+2r_h)(a^2+b^2) \\
	\nonumber a_1&=& 8a(a^2+3b^2)\\
	\nonumber a_2&=&32 a b\\
	\nonumber a_3&=&16 a.
\eea
The corresponding horizon values read,
\bea
	\nonumber\hat{\omega}_\mathrm{h}   &=& -\frac{a_0 - 8a^3 r_\mathrm{h} \sin^2\theta}{(a^2+b^2)^3}\\[5mm]
	\nonumber\hat{\Phi}_\mathrm{h}     &=& 2 Q\,\frac{a^4 - b^4 + 4 a^2 b r_\mathrm{h} \sin^2 \theta}{(a^2+b^2)^3}.
\eea
We thus obtain in the limit $r_\mathrm{h}= 0$:
\bea
	\nonumber\hat{\mu}_\mathrm{h} &=& M^2+a^2 \cos^2\theta \\
	\nonumber\hat{u}_\mathrm{h}   &=& \frac{\left(a^2+M^2\right)^2}{M^2+a^2 \cos ^2\theta}\\
	\nonumber\omega_\mathrm{h}    &=& \frac{a}{a^2+M^2}\\
	\nonumber\Phi_\mathrm{h}      &=& \frac{QM}{a^2+M^2}\\
	\nonumber(A_\varphi)_\mathrm{h} &=& -\frac{Q a M \sin ^2\theta}{M^2+a^2 \cos^2\theta}\\
	\nonumber\hat{\omega}_\mathrm{h}&=&-\frac{4 a M}{\left(a^2+M^2\right)^2}\\
	\nonumber\hat{\Phi}_\mathrm{h}  &=&- \frac{2 Q^3}{\left(a^2+M^2\right)^2}.
\eea
Note that for the entire Kerr-Newman family the inequality (\ref{in:JQA}) holds, with $p_J$ and $p_Q$ defined in (\ref{p_J}, \ref{p_Q}). The equal sign is assumed if and only if the black hole is degenerate, $\kappa=0$.
\section{Nonlinear study of the system (\ref{Eqn_for_ud_2}, \ref{Eqn_for_Aphi_2})}\label{NonLinSystem}

Defining the dimensionless functions
\beq\label{Def_f_and_g}
	f = \frac{\hat{\omega}_\mathrm{h}}{\hat{B}_\mathrm{h}}\hat{u}\;,\qquad
	g = \frac{1-\alpha^2}{1+\alpha^2}\,\hat{\Phi}_\mathrm{h}^{-1}
		\left[\hat{\omega}_\mathrm{h}A_\varphi-\hat{\Phi}_\mathrm{h}\right],
			\;,
\eeq
we can eliminate the constants $\hat{B}_\mathrm{h}, \hat{\omega}_\mathrm{h}$ and $\hat{\Phi}_\mathrm{h}$ in (\ref{Eqn_for_ud_2}, \ref{Eqn_for_Aphi_2}). Furthermore we restrict ourselves to equatorially symmetric solutions of the system (\ref{Eqn_for_ud_2}, \ref{Eqn_for_Aphi_2}). It is therefore appropriate to transform to the variable $x = \cos^2\theta$ which leads to the following differential equations:
\bea
	\lefteqn{\nonumber 2x(1-x)f''+(1-3x)f'-f+\frac{1}{2}(1-x)f^3-2x(1-x)\frac{f'^2}{f}}\\
	\label{Eqn_for_f}\hspace*{4cm} &=&-\frac{1-\alpha^2}{2\alpha}\left(4xg'^2+f^2g^2\right)
	\\[5mm]
	\label{Eqn_for_g} 
	2xg''+g' &=& 2x\frac{g'f'}{f}-\frac{1}{2}gf^2
\eea
where the prime $(')$ denotes the derivative with respect to $x$.
Since $f(x)$ and $g(x)$ have to be regular near $x=0$, there have to exist convergent power series expansions:
\beq\label{SumAnsatz}
f(x) = \sum_{n=0}^{\infty}c_nx^n \qquad g(x) = \sum_{n=0}^{\infty}d_nx^n.
\eeq
Insertion of (\ref{SumAnsatz}) into (\ref{Eqn_for_f}, \ref{Eqn_for_g}) allows us to calculate the coefficients $c_n$ and $d_n$ for $n \geq 1$ explicitly as functions of $c_0$ and $d_0$, which
serve as the two integration constants remaining due to the equatorial symmetry being imposed. 
In more detail, with the parameters $c:=c_0^2/4\geq 0$ and $d:=c_0d_0^2$, the quotients $\tilde{c}_n = c_n/c_0$ and
$\tilde{d}_n = d_n/d_0$ for $n\geq 1$ allow the expansions:
\bea
\label{tilde_c} \tilde{c}_n &=& \sum_{k=0}^n d^k \sum_{l=0}^{n-k} p^{(n)}_{k,l}c^l \\
\label{tilde_d} \tilde{d}_n &=& \sum_{k=0}^{n-1} d^k \sum_{l=1}^{n-k} q^{(n)}_{k,l}c^l
\eea
with numerical coefficients $p^{(n)}_{k,l}, q^{(n)}_{k,l}$. An algebraic
computer program may allow one to produce explicit expressions for
$p^{(n)}_{k,l}$ and $q^{(n)}_{k,l}$ for all $n$. The conjecture that the system
(\ref{Eqn_for_f}, \ref{Eqn_for_g}) permits only the regular solutions (\ref{uh_2}, \ref{Aphih_2}) is then equivalent to the condition that at least one of the series (\ref{SumAnsatz}) diverges at $x=1\;(\sin\theta =0)$ for all parameter values $c$ and $d$, except for the combination
\[c +\frac{1-\alpha^2}{2\alpha}d - 1= 0. \]
This conjecture is supported by the fact that, due to the explicit form of (\ref{uh_2}, \ref{Aphih_2}), the coefficients have the structure
\bea
	\nonumber \tilde{c}_n &=& \left(c +\frac{1-\alpha^2}{2\alpha}d - 1\right)\hat{c}_n+(-1)^nc^n \\
	\nonumber \tilde{d}_n &=& 2c\left[\left(c +\frac{1-\alpha^2}{2\alpha}d - 1\right)\hat{d}_n+(-1)^nc^{n-1}\right]
\eea
with new coefficients $\hat{c}_n$ and $\hat{d}_n$. With $c_0=2\alpha$ and
$d_0=-1$, the latter terms of $\tilde{c}_n$ and
$\tilde{d}_n$ sum up to the solutions (\ref{uh_2}, \ref{Aphih_2}):
\bea
	\nonumber f_0(x)&=&\frac{2\alpha}{1+\alpha^2x}\\
	\nonumber g_0(x)&=&-\frac{1-\alpha^2x}{1+\alpha^2x}.
\eea
It remains to prove that $\sum\limits_{n=1}^{\infty}\hat{c}_n x^n$ and/or
$\sum\limits_{n=1}^{\infty}\hat{d}_n x^n$ diverge for $x=1$.\\

It may well be possible that a complete analysis of the system (\ref{Eqn_for_ud_2}, \ref{Eqn_for_Aphi_2}) can be carried out within the framework of the `inverse scattering theory' which allows one to associate the nonlinear Einstein-Maxwell equations with an appropriate linear matrix problem \cite{NeugebKram85}. However, such a study is beyond the scope of this paper.

\section{Linear study of the system (\ref{Eqn_for_ud_2}, \ref{Eqn_for_Aphi_2})}\label{Annex_to_deg}

We prove that the solution (\ref{uh_2}, \ref{Aphih_2}) of the nonlinear, coupled
system of Einstein-Maxwell equations (\ref{Eqn_for_ud_2}, \ref{Eqn_for_Aphi_2}) is an isolated solution, i.e. that
there exist no neighbouring solutions to (\ref{uh_2}, \ref{Aphih_2}) fulfilling the geometrical
and physical conditions of regularity, equatorial symmetry, and axis regularity.
In order to simplify the analysis, we introduce here the variable $z = 1 + \alpha^2
\cos^2\theta$. For the neighbouring solutions we make the following ans\"atze for $f$ and $g$ (see \ref{NonLinSystem}, eqn. (\ref{Def_f_and_g})):
\bea
\label{a:1}
f(z) &=& \frac{2\alpha}{z} + \epsilon k(z)\\
\label{a:2}
g(z) &=& -\frac{2-z}{z} + \frac{\epsilon}{\alpha(1-\alpha^2)}\; h(z).
\eea
We consider the corresponding differential equations for $k(z)$ and $h(z)$ to first order in $\epsilon$. They read (with the prime now denoting the $z$-derivative):
\bea
\nonumber\fl 2z^2(z-1)(z-1-\alpha^2)k'' + z\left[7z^2-(11+5\alpha^2)z+4(1+\alpha^2)\right]k' &&\\
\fl\nonumber \quad +\;\left[3z^2+2(1-\alpha^2)z-4(1+\alpha^2)-\frac{2}{z}(1-\alpha^2)(z-2)^2\right]k &&\\[2mm] 
\fl\label{a:3} \quad = \;4\,\left[2(z-1)h'+\left(1-\frac{2}{z}\right)\,h\right] &&\\ \nonumber 
\eea

\beq
\fl\label{a:4}  4\left[2(z-1)h'' + \frac{3z-2}{z}h' + \frac{2}{z^2}\,h\right]  =
	 8\,(1-\alpha^2)\left[\left(1-\frac{1}{z}\right)k' + \frac{1}{z^2}\,k\right].
\eeq

Surprisingly, it turns out that, due to the very special structure of the
Einstein-Maxwell equations, the $z$-derivative of the right hand side of (\ref{a:3}) equals the left hand side of (\ref{a:4}). Therefore, the $z$-derivative of (\ref{a:3}) results in a separate,
homogeneous differential equation of third order for the function $k(z)$, in
which, furthermore, the terms with factors $z^{-1}$ and $z^{-2}$ cancel:
\bea
\nonumber 2z(z-1)(z-1-\alpha^2)k''' + [15z^2-(23+11\alpha^2)z+8(1+\alpha^2)]k''&&\\
\nonumber \quad + \left[24z-(22+10\alpha^2)\right]k' + 6\,k = 0.
\eea
And even more remarkably, this equation can be explicitly integrated twice:
\beq\label{a:6}
2z(z-1)k' + (3z-4)k = c_1 + \frac{c_2}{1+\alpha^2-z}
\eeq
with arbitrary constants $c_1, c_2.$ \\
Equation (\ref{a:6}) can now be explicitly integrated by standard methods (solution of
the homogeneous equation, and variation of the integration constant). If we
transform back from $z$ to the angular variable $\theta$, we get (with an
additional constant $c_3$):
\bea
\nonumber
k(\theta) &=& c_3\frac{\cos\theta}{(1+\alpha^2\cos^2\theta)^2} -
c_1\frac{1-\alpha^2\cos^2\theta}{(1+\alpha^2\cos^2\theta)^2}  \nonumber\\ && -
\nonumber
\frac{c_2}{\alpha^2(1+\alpha^2\cos^2\theta)^2} \left[1+\frac{1+\alpha^2}{2}\cos\theta
\left(\ln\frac{1-\cos\theta}{1+\cos\theta}\right)\right].
\eea

Since the first term is not equatorially symmetric, and the last term is
singular at $\theta=0$ and $\theta=\pi$, the constants $c_2$ and $c_3$
have to be zero. Insertion of the remaining term into (\ref{a:3}) leads to the
differential equation
\[ 2z(z-1)h' + (z-2)h = -c_1 \left[\frac{3-5\alpha^2}{4}z -
\frac{2(1-\alpha^2)}{z^2}(2z^2-3z+2)\right]
\]
with the general solution (in the variable $\theta$)
\bea\nonumber
h(\theta) &=& c_4\frac{\cos\theta}{1+\alpha^2\cos^2\theta} \\ && \label{a:9}
- c_1 \left[\frac{3-5\alpha^2}{4} +
\frac{5-3\alpha^2}{2(1+\alpha^2\cos^2\theta)} -
\frac{2(1-\alpha^2)}{(1+\alpha^2\cos^2\theta)^2}\right]
\eea
where $c_4$ is an arbitrary integration constant.\\
Now the condition of equatorial symmetry of the solution $h(\theta)$ enforces
the constant $c_4$ to be zero. The second term of (\ref{a:9}) fulfils the axis
condition $h(\theta=0)=0$ only in the uncharged case $\alpha^2=1$. 
However, in this case $h(\theta)$ has to be identically zero in order to realize the regularity of $A_\varphi$, see (\ref{a:2}) and (\ref{Def_f_and_g}).
Therefore, we have for
all cases $c_1=0$ which completes the proof that there are no physically
admissible, neighbouring solutions to the solutions (\ref{Eqn_for_ud_2}, \ref{Eqn_for_Aphi_2}).

\section*{References}
\bibliographystyle{unsrt}
\bibliography{Reflink}

\end{document}